\documentclass[a4paper,10pt]{iopart}
\usepackage[numbers,square,sort&compress]{natbib}

\usepackage{graphicx}
\usepackage{bm}
\usepackage{xcolor}

\newcommand{\onlinecite}[1]{\cite{#1}}

\newcommand{\mbf}[1]{\mathbf{#1}}
\newcommand{\ddt}[0]{\frac{\partial}{\partial t}}

\renewcommand{\t}[1]{\textrm{#1}}
\renewcommand{\k}[0]{\mbf{k}}

\newcommand{\up}[0]{\uparrow}
\newcommand{\down}[0]{\downarrow}
\newcommand{\nn}[0]{\nonumber\\}

\newcommand{\C}[3]{C_{l#1\k#3}^{l#2}}

\newcommand{\bstd}[0]{{b_{\beta\k_1}^{\alpha\k_2}}}

\newcommand{\Jsd}[0]{J_{sd}}

\newcommand{\nMn}[0]{n_{Mn}}
\newcommand{\NMn}[0]{N_{Mn}}

\begin{document}
\title[]{Effective Equations for the Precession Dynamics of Electron Spins and 
Electron-Impurity Correlations in Diluted Magnetic Semiconductors}
\author{M.~Cygorek, V.~M.~Axt}
\address{ Theoretische Physik III, Universit{\"a}t Bayreuth, 95440 Bayreuth, Germany}
\begin{abstract}
Starting from a quantum kinetic theory for the spin dynamics in diluted
magnetic semiconductors,  we derive simplified  equations that effectively describe the 
spin transfer between carriers and magnetic impurities for an arbitrary
initial impurity magnetization.
Taking the Markov limit of these effective equations, we obtain 
good quantitative agreement with the full quantum kinetic theory 
for the spin dynamics in bulk systems at high magnetic doping. 
In contrast, 
the standard rate description where the carrier-dopant interaction 
is treated according to Fermi's golden rule,
which involves the assumption of a short memory as well as 
a perturbative argument,
has been shown previously to fail if the impurity magnetization is non-zero.
The Markov limit of the effective equations is derived, assuming only 
a short memory, while higher order terms are still accounted for.
These higher order terms
represent the precession of the
carrier-dopant correlations in the effective magnetic field due to the impurity spins.
Numerical calculations show that the Markov limit of our effective equations 
reproduces the results of the full quantum kinetic theory very well.
Furthermore, this limit allows for analytical solutions and 
for a physically transparent interpretation.
\end{abstract}
\pacs{75.78.Jp, 75.50.Pp, 75.30.Hx, 72.10.Fk}
\maketitle
\section{Introduction}
Diluted magnetic semiconductors (DMS), in particular Mn doped II-VI and 
III-V materials, have been studied for several 
decades\cite{KossutRate3D,FurdynaReview,Krenn,Crooker97,
Awschalom_RSA,Akimov,Roennburg,Cywinski,Wu09,Perakis_Wang,WuReview,intro,
Camilleri01,FewMn08, HyperfineCdMnTe, ExcSpinDMSPhonons,DMS_Superlattices,
PulsedB_DMS}.
However, the theoretical description of the ultrafast spin dynamics of
the magnetic impurities and carriers is, so far, mostly limited to a single-particle mean-field
picture, where transfer rates are calculated perturbatively by Fermi's golden rule.
Interesting features of the spin dynamics in DMS that have been 
demonstrated in recent time-resolved Kerr measurements\cite{BenCheikh2013},
like the nonmonotonous dependence of
the transverse spin dephasing time in extremely diluted Cd$_{1-x}$Mn$_x$Te quantum wells
or the mismatch between the theoretically predicted and experimentally measured dephasing
times for zero magnetic field, still lack a satisfactory theoretical
explanation.
To provide a more elaborate theoretical framework for the discussion and 
quantitative calculation of the spin dynamics in DMS, 
a quantum kinetic theory based on a correlation expansion has
been introduced\cite{Thurn:12} where the exchange interaction between free carriers
and the $d$ electrons of the Mn impurities was modelled by a Kondo Hamiltonian.
The full quantum kinetic theory is, however, numerically challenging and the physical interpretation 
requires some effort. Hence, it is a difficult task to efficiently implement other 
mechanisms of spin exchange and dephasing into the theory in order 
to account for effects that are in many cases needed for a proper 
description of real experiments like, e.~g.,
the D'yakonov-Perel'\cite{DP}, Elliot-Yafet\cite{EY1,EY2} and 
Bir-Aronov-Pikus\cite{BAP} mechanisms. 
However, it was already shown that for three dimensional systems in which the 
number of Mn impurities $\NMn$ exceeds the number of free carriers $N_e$,
a simplification of the quantum kinetic theory 
can be established 
that reasonably reproduces 
results in the case of a vanishing initial Mn magnetization\cite{Thurn:13_1}. 
This was achieved by a perturbative treatment of the carrier-impurity
interaction 
as well as the assumption of a short memory.
This procedure yielded the same rate equations as Fermi's golden rule.
In contrast, for a nonzero average Mn spin these
rate  equations were shown to describe
only the electron spin component perpendicular to the Mn spin well,
while a discrepancy in the dynamics of the parallel electron spin component 
could be attributed to neglected terms of higher than leading order in 
the coupling constant $\Jsd$ of the Kondo Hamiltonian (\ref{eq:kondoHam}) 
in the perturbative derivation of the 
rate equations in Ref.~\onlinecite{Cygorek:14_1}.

In the present article, we derive approximate equations of motion for the electron 
spins in the spirit of the 
equations in Ref.~\onlinecite{Cygorek:14_1}, but take 
the higher order corrections into account. These equations describe the  
effects of the precession of the electron spins 
around the effective magnetic field due to the Mn magnetization 
and effectively account for a precession-type dynamics of the electron-Mn correlations that 
has been identified previously in Ref.~\onlinecite{Cygorek:14_1}.
The resulting 
{\em precession of electron spins and correlations} (PESC) equations  
are then discussed and their Markov limit
is established which can be solved analytically. Numerical calculations
show that for $\NMn\gg N_e$ these analytical solutions coincide with the results
of the full quantum kinetic theory, at least in three dimensional systems.
The simplicity of the PESC equations makes it possible to easily interpret
the basic physical processes involved in the quantum kinetic theory and allows the PESC 
equations to provide a suitable framework
for further studies of non-Markovian effects as well as
of the interplay between the $s$-$d$ interaction and 
other mechanisms of spin relaxation and dephasing.

It is noteworthy that the derived effective equations are expected to be applicable not only for 
the spin dynamics in DMS, but they can easily be extended to describe more generally any system, 
in which a continuum of states is coupled to localized magnetic impurities via a Kondo-like Hamiltonian
\begin{eqnarray}
H=\sum_{l\k}E_{\k}c^\dagger_{l\k}c_{l\k}+
\Jsd\sum_{Ii}\hat{\mbf S}^I\cdot \hat{\mbf s}^i \delta(\mbf R_I-\mbf r_i),
\label{eq:kondoHam}
\end{eqnarray}
where in the case of DMS $c^\dagger_{l\k}$ and $c_{l\k}$ describe creation and annihilation 
operators for conduction band electrons with wave vector $\k$ in the subband $l$, 
$E_{\k}$ are the corresponding single-particle energies, $\hat{\mbf S}^I$ and $\hat{\mbf s}^i$
are the spin operators while $\mbf R_I$ and $\mbf r_i$ are the positions
of the $I$-th Mn ion and the $i$-th electron, respectively.
Similar magnetic interactions can also arise from nuclear spins due to the 
Fermi contact interaction 
or an effective interaction between conduction
band electrons and localized states, such as in quantum dots, or quasi-particles, e.~g. excitons, 
in a huge variety of systems
ranging from semiconductor heterostructures to novel materials such as graphene or dichalcogenides, since
the main difference between these systems lies in the details of the single-particle band structures.
Therefore, the equations of motion studied here are of prototypical character for 
the spin dynamics of extended systems.

The article is outlined as follows: First, we summarize the quantum kinetic theory and 
reproduce the basic equations of motion where we restrict ourselves to the
terms that were shown in Ref.~\onlinecite{Cygorek:14_1}
to be numerically important in the case $\NMn\gg N_e$.
In a next step, we apply
a rotating-wave-like approximation and derive the PESC equations of motion for 
the electron spins and occupations.
Then, the Markov limit of the PESC equations is introduced, 
and the thereby described physical effects are discussed; in particular the
spectral redistribution of electrons as well as Pauli blocking effects
are shown to arise naturally on this level of theory. Subsequently, analytical
solutions to the Markov limit of the PESC equations are presented and compared with 
numerical results of the full quantum kinetic theory. 

\section{Method: Derivation of Effective Equations}
We will give a short overview of the quantum kinetic theory 
for the spin dynamics in DMS developed in Ref.~\onlinecite{Thurn:12}. There,
a systematic derivation of equations of motion for the spins of interacting 
carriers and Mn impurities in DMS has been presented accounting for conduction and valance
band carriers, their coherences, the Mn impurity spins, the correlations between 
carriers and impurities as well as the effect of an external laser field, where
a disorder average over the random distribution of the impurities in the semiconductor
was performed. 
Apart from the corresponding band energies, the theory accounts for
the exchange interaction between carriers and Mn impurities
as well as for the dipole coupling to a classical laser field. 

We want to focus our study on the spin dynamics in isoelectrically doped bulk 
DMS starting from a non-equilibrium state. Such kind of situation 
can be prepared, e.g., by
optical excitation with circularly polarized light.
Since in bulk systems, the typical timescale of the hole spin relaxation is
of the order of $100$ fs\cite{HoleSpinBulkGaAs} due to the strong spin-orbit interaction, 
we can neglect the valence band and the interband coherences when concentrating on a ps timescale. 
When only the conduction band electrons and the 
impurities together with their correlations are considered,
the resulting equations of motion can be simplified as it was shown in 
Ref.~\onlinecite{Cygorek:14_1}. 
The dynamical variables
used in this subset of equations are the conduction band electron density matrices
$C_{l_1\k_1}^{l_2}$, the average Mn density matrix $M_{n_1}^{n_2}$ and their
correlations $Q_{l_1n_1\k_1}^{l_2n_2\k_2}$:
\numparts
\begin{eqnarray}
C_{l_1\k_1}^{l_2}&=\langle c^\dagger_{l_1\k_1}c_{l_2\k_1}\rangle, \\
M_{n_1}^{n_2}&=\langle \hat{P}^I_{n_1n_2} \rangle, \\
Q_{l_1n_1\k_1}^{l_2n_2\k_2}&=\delta\langle c^\dagger_{l_1k_1}c_{l_2\k_2}
\hat{P}^I_{n_1n_2}e^{i(\k_2-\k_1)\mbf R_I}\rangle \nn&+
\delta\langle c^\dagger_{l_1k_1}c_{l_2\k_2}e^{i(\k_2-\k_1)\mbf R_I}\rangle
\langle \hat{P}^I_{n_1n_2} \rangle,
\end{eqnarray}
\endnumparts
where $c^\dagger_{l_1\k_1}$ and $c_{l_1\k_1}$ are the electron creation and 
annihilation operators, 
$\hat{P}^I_{n_1n_2}=|I,n_1\rangle\langle I,n_2|$ is the density operator
for the spin-$\frac 52$ state of the $d$ electrons of the $I$-th Mn ion,
and the indices $l_i\in\{1,2\}$ 
as well as $n_i\in\big\{-\frac 52,-\frac 32,\dots,\frac 52\big\}$ 
represent spin indices of the conduction band electrons and Mn spin states, 
respectively.
The brackets symbolize the quantum mechanical average as well as the disorder 
average over the random distribution of the Mn positions $\mbf R_I$ and
$\delta\langle\dots\rangle$ describes the true correlations, i.~e., the average value
minus all possible factorized parts
(cf. Refs.~\onlinecite{Thurn:12} for the details of the correlation expansion and
the truncation scheme).

In Ref.~\onlinecite{Cygorek:14_1} it was also found that considering only the 
numerically important terms, the correlations can be summed up together
with the electron and Mn spin matrices yielding new correlation functions 
with fewer degrees of freedom:
\begin{eqnarray}
Q^{\alpha \k_2}_{\beta\k_1}:= \sum_{\stackrel{l_1l_2}{n_1n_2}}S^\beta_{n_1n_2}
s^\alpha_{l_1l_2} Q^{l_2n_2\k_2}_{l_1n_1\k_1},
\end{eqnarray}
where $s^\alpha_{l_1l_2}$ with $\alpha \in \{0,1,2,3\}$
are the electron spin matrices and the identity matrix (for $\alpha=0$), 
respectively, and $S^\beta$ with $\beta \in \{1,2,3\}$ are the 
spin-$\frac 52$
matrices for the Mn ions.
The equations of motion for the electron occupations $n_{\k_1}=\sum_{l}C^l_{l\k_1}$ 
and spins $s^\alpha_{\k_1}=\sum_{ll'}s^\alpha_{ll'}C_{l\k_1}^{l'}$ 
are then given by\cite{Thurn:12,Cygorek:14_1}:
\numparts
\begin{eqnarray}
\label{eq:dgln}
\label{eq:grundglA}
\ddt n_{\k_1}&=\frac \Jsd\hbar \nMn \frac 1V\sum_\k 
\sum_{i=1}^3 
2\Im\big(
Q^{i\k}_{i\k_1}\big),\\
\label{eq:grundglB}
\ddt s_{\k_1}^\alpha&=\frac \Jsd\hbar \nMn\bigg\{ \big(\langle \mbf S\rangle\times
\mbf s_{\k_1}\big)_\alpha +\nn&+
\frac 1V\sum_\k\bigg[\frac 12\Im\big(Q^{0\k}_{\alpha\k_1}\big)
+\sum_{ij=1}^3 
\epsilon_{ij\alpha}\Re\big(Q^{j\k}_{i\k_1}\big)\bigg]\bigg\},\\
\label{eq:grundglC}
\ddt Q^{\alpha\k_2}_{\beta\k_1}&=-i(\omega_{\k_2}-\omega_{\k_1})
Q^{\alpha\k_2}_{\beta\k_1}+\nn&+
\frac i\hbar\Jsd \left(
{b^{\alpha\k_2}_{\beta\k_1}}^{I.1}+{b^{\alpha\k_2}_{\beta\k_1}}^{I.2}
+{b^{\alpha\k_2}_{\beta\k_1}}^{II.1}
+{b^{\alpha\k_2}_{\beta\k_1}}^{Res}\right),
\label{eq:dglQ}
\end{eqnarray}
with source terms
\begin{eqnarray}
{b^{\alpha\k_2}_{\beta\k_1}}^{I.1}&=\sum_\gamma \sum_{ll_1l_2}\Big(
\langle S^\gamma S^\beta\rangle s^\gamma_{ll_1} s^\alpha_{l_1l_2}\C{}{_2}{_2}
-\langle S^\beta S^\gamma \rangle s^\alpha_{l_1l_2}s^\gamma_{l_2l}\C{_1}{}{_1}
\Big),\\
{b^{\alpha\k_2}_{\beta\k_1}}^{I.2}&=-i \sum_{\gamma\delta}
\epsilon_{\gamma\beta\delta}\sum_{ll'l_1l_2}
\langle S^\delta\rangle \big\{
s^\gamma_{ll'}\C{_1}{'}{_1} s^\alpha_{l_1l_2}\C{}{_2}{_2}\big\},\\
{b^{\alpha\k_2}_{\beta\k_1}}^{II.1}&= i\sum_{\gamma\delta}
\epsilon_{\gamma\alpha\delta} \nMn
\langle S^\gamma\rangle Q^{\delta\k_2}_{\beta\k_1}
=-i\frac \hbar\Jsd \big(\boldsymbol{\omega}_M \times 
{\mbf Q_{\beta\k_1}^{\phantom{\beta}\k_2}}\big)_\alpha,
\label{eq:grundglF}
\end{eqnarray}
\endnumparts
where $\langle S^\alpha \rangle=\sum\limits_{n_1n_2}S^\alpha_{n_1n_2}M_{n_1}^{n_2}$ are the averages
of the Mn spin operators and 
$\langle S^\alpha S^\beta\rangle=\sum\limits_{n_1n_2n}S^\alpha_{n_1n}S^\beta_{nn_2}M_{n_1}^{n_2}$ 
its second moments.
$V$ represents the volume of the sample, $\nMn=\frac{\NMn}V$ is the Mn density and
the $\k$-sum has to be performed over all states in the first Brillouin zone.
${b^{\alpha\k_2}_{\beta\k_1}}^{Res}$ comprise the residual source terms that 
were identified in Ref.~\onlinecite{Cygorek:14_1} to be insignificant if 
$\NMn\gg N_e$ and $V$ is large. Therefore, we will henceforth neglect 
${b^{\alpha\k_2}_{\beta\k_1}}^{Res}$. 

In Eq.~(\ref{eq:grundglF}), the mean field precession frequency and axis of the 
electrons around the Mn magnetization $\boldsymbol{\omega}_M:=\frac \Jsd\hbar\nMn
\langle \mbf S\rangle$ has been introduced as well as the vector representation of 
the correlations $\big(\mbf  Q_{\beta\k_1}^{\phantom{\beta}\k_2}\big)_\alpha:=
Q_{\beta\k_1}^{\alpha\k_2}$.
$\omega_{\k}=\frac{E_{\k}}\hbar=\frac{\hbar \k^2}{2m^*}$ describes the 
single-particle frequencies of the quasi-free conduction band electrons assuming
a parabolic band structure with effective mass $m^*$ without the electron-Mn 
exchange interaction.
${b^{\alpha\k_2}_{\beta\k_1}}^{I.1}$ is the source term describing the 
build-up of correlations between the impurities and the carriers, while
${b^{\alpha\k_2}_{\beta\k_1}}^{I.2}$ provides corrections to 
${b^{\alpha\k_2}_{\beta\k_1}}^{I.1}$ due to (mean field)
two-electron effects, e.~g., Pauli blocking\cite{Cygorek:14_1}.
The precession-type dynamics of the carrier-Mn correlations 
around the effective magnetic field due 
to the Mn magnetization 
are incorporated via the term ${b^{\alpha\k_2}_{\beta\k_1}}^{II.1}$.
The neglect of the latter has been found to be the reason for the failure
of the golden rule-type rate equations of Ref.~\onlinecite{Cygorek:14_1}
in describing the parallel spin transfer 
between the carriers and the magnetic impurities. 

It is, however, possible to account for this precession and to integrate
Eq.~(\ref{eq:grundglC}) formally. This is particularly easy if we
use the assumption $N_{Mn}\gg N_e$ that allows us to regard the 
Mn density matrix as nearly constant in time. If the z-axis is defined to point in
the direction of the Mn magnetization, the correlations are given by:
\numparts
\begin{eqnarray}
\label{eq:QmitGedA}
Q_{\beta\k_1}^{0\k_2}&=\frac i\hbar \Jsd
\int\limits_0^tdt'\Big\{ {b_{\beta\k_1}^{0\k_2}}^I(t')
e^{i(\omega_{\k_2}-\omega_{\k_1})(t'-t)}\Big\},\\
Q_{\beta\k_1}^{x\k_2}&=\frac i{2\hbar} \Jsd
 \int\limits_0^tdt'\Big\{ 
\big({b_{\beta\k_1}^{xk_2}}^I(t')+i {b_{\beta\k_1}^{y\k_2}}^I(t')\big)
e^{i(\omega_{\k_2}-\omega_{\k_1}-\omega_M)(t'-t)}+\nn&\qquad\qquad
+\big({b_{\beta\k_1}^{x\k_2}}^I(t')-i {b_{\beta\k_1}^{y\k_2}}^I(t')\big)
e^{i(\omega_{\k_2}-\omega_{\k_1}+\omega_M)(t'-t)}\Big\}\\
Q_{\beta\k_y}^{y\k_2}&=\frac i{2\hbar} \Jsd \int\limits_0^tdt'\Big\{
\big({b_{\beta\k_1}^{y\k_2}}^I(t')-i{b_{\beta\k_1}^{x\k_2}}^I(t')\big)
e^{i(\omega_{\k_2}-\omega_{\k_1}-\omega_M)(t'-t)}+\nn&\qquad\qquad
+\big({b_{\beta\k_1}^{y\k_2}}^I(t')+i {b_{\beta\k_1}^{x\k_2}}^I(t')\big)
e^{i(\omega_{\k_2}-\omega_{\k_1}+\omega_M)(t'-t)}\Big\}\\
Q_{\beta\k_1}^{z\k_2}&=\frac i{\hbar} \Jsd\int\limits_0^tdt'\Big\{ {b_{\beta\k_1}^{z\k_2}}^I(t')
e^{i(\omega_{\k_2}-\omega_{\k_1})(t'-t)}\Big\}
\end{eqnarray}
\endnumparts
where ${b_{\beta\k_1}^{\alpha\k_2}}^I:={b_{\beta\k_1}^{\alpha\k_2}}^{I.1}+
{b_{\beta\k_1}^{\alpha\k_2}}^{I.2}$.
In order to simplify Eqs.~(5), 
we follow the line of Ref.~\onlinecite{Cygorek:14_1} 
and identify fast and slowly changing terms.
To this end, we express the electron spin in the state with k-vector $\k_1$ 
\begin{eqnarray}
\mbf s_{\k_1}:=\left(\begin{array}{c}
s^\perp_{\k_1} \cos(\omega_M t +\varphi_{\k_1}) \\
s^\perp_{\k_1} \sin(\omega_M t +\varphi_{\k_1}) \\
s^\|_{\k_1}
\end{array}\right),
\end{eqnarray}
in terms of the spin component parallel to the Mn magnetization $s^\|_{\k_1}$,
the perpendicular spin component $s^\perp_{\k_1}$ and the phase $\varphi_{\k_1}$.
A rotating-wave-like approximation is established, by assuming that these variables
$s^\|_{\k_1}$,$s^\perp_{\k_1}$ and $\varphi_{\k_1}$ as well as the 
electron occupation $n_{\k_1}$ of the states with k-vector $\k_1$ change
only slowly in time,
since they are constant in the mean field approximation.
When they are drawn out of the integrals in 
Eqs.~(5)  
and the resulting expressions
for the correlations are inserted in the equations of motion 
(\ref{eq:grundglA}) and (\ref{eq:grundglB}) for the
electron occupations and spins, we get:
\numparts
\begin{eqnarray}
\label{eq:modmarkovA}
\ddt n^{\up/\down}_{\k_1}&=\sum_\k\bigg\{\Re(G_{\omega_\k}^{\omega_{\k_1}})
\frac{b^\|}2 \left[n_\k^{\up/\down}-n_{\k_1}^{\up/\down}\right]+\nn&+
\Re(G_{\omega_\k}^{\omega_{\k_1}\pm\omega_M})
\left[b^\pm n^{\down/\up}_\k-b^\mp  n^{\up/\down}_{\k_1}
\mp 2 b^0 n^{\up/\down}_{\k_1}n^{\down/\up}_\k \right]\bigg\}\\
\ddt \mbf s^\perp_{\k_1}&=-\sum_\k\bigg[
\Re(G_{\omega_\k}^{\omega_{\k_1}-\omega_M})\big(\frac{b^+}2-b^0n^\up_\k\big)\mbf s^\perp_{\k_1}+
\nn&+
\Re(G_{\omega_\k}^{\omega_{\k_1}+\omega_M})\big(\frac{b^-}2+b^0n^\down_\k\big)\mbf s^\perp_{\k_1}+
\Re(G_{\omega_\k}^{\omega_{\k_1}})\frac{b^\|}2\big(\mbf s^\perp_\k+\mbf s^\perp_{\k_1}\big)\bigg]+
\nn&+
\frac{\langle \mbf S\rangle}{|\langle \mbf S\rangle|}\times \bigg[\omega_M
-\sum_\k\Big\{\Im(G_{\omega_\k}^{\omega_{\k_1}-\omega_M})\big(\frac{b^+}2-b^0n^\up_\k\big)
+\nn&
-\Im(G_{\omega_\k}^{\omega_{\k_1}+\omega_M})\big(\frac{b^-}2+b^0n^\down_\k\big)\Big\}
\bigg] \mbf s_{\k_1}^\perp,
\label{eq:modmarkovB}
\end{eqnarray}
\endnumparts
where in favor of a compact notation, the variables for the occupations and spins
have been transformed into the occupations of the spin-up and spin-down band,
i.~e., the diagonal elements of the reduced electron density matrix,
according to:
\begin{eqnarray}
n^{\up/\down}_{\k_1}:=\frac{n_{\k_1}}2 \pm s^\|_{\k_1}.
\end{eqnarray}
The coefficients used in Eqs.~(7)  
are given by
$b^{\pm}:=\langle {S^\perp}^2\rangle\pm \frac{\langle S^\|\rangle}2$,
$b^{0}:=\frac{\langle S^\|\rangle}2$
as well as 
$b^{\|}:={\langle {S^\|}^2 \rangle}$,
where $S^\|:= \hat{\mbf S} \cdot \frac{\langle \hat{\mbf S}\rangle}{
|\langle \hat{\mbf S}\rangle|}$ is the Mn spin operator component parallel to
the average Mn spin and 
$\langle {S^\perp}^2\rangle=\frac 12\langle S^2-{S^\|}^2 \rangle$.
The remaining integral together with some prefactors are subsumed into the 
memory function 
\begin{eqnarray}
G_{\omega_\k}^{\omega_{\k_1}}&:=\frac{\Jsd^2}{\hbar^2}\frac{\nMn}V\int\limits_{-t}^0 dt' 
e^{i(\omega_\k-\omega_{\k_1})t'}.
\label{eq:memory}
\end{eqnarray}
Eqs.~(7) 
together with the memory in Eq.~(\ref{eq:memory})
describe the spin dynamics of the conduction band electron, where the precession
of the electron spins and electron-impurity correlations are accounted for
and will henceforth be referred to as the PESC equations.
Note that to account for finite memory effects, 
the memory $G_{\omega_\k}^{\omega_{\k_1}}$ has to 
be regarded as an integral operator and the spins and occupations in the 
r.h.s. of Eqs.~ (7) 
have to be evaluated at the time $t+t'$.

\section{Results and Discussion}
\subsection{Markov Limit of Effective Equations}
The Eqs.~(7) 
are written in terms of dynamical variables 
that depend on the $k$-vector including the angles. This is important 
for possible extensions of the theory with $k$-dependent 
effective magnetic fields resulting from, e.~g., Dresselhaus\cite{Dresselhaus}- and 
Rashba\cite{Rashba}-terms.
Without such extensions, angle-averaged equations can be obtained after going 
over to the Markov limit from which the physical meaning of the 
individual terms in the PESC Eqs.~(7) 
will become most obvious. Technically, this 
is done by letting 
the lower integral bound $-t$ go to $-\infty$ in the memory function
$G_{\omega_\k}^{\omega_{\k_1}}$.
The memory is then given by:
\begin{eqnarray}
G_{\omega_\k}^{\omega_{\k_1}}&
\approx\frac{\Jsd^2}{\hbar^2}\frac{\nMn}V\left\{\pi\delta(\omega_\k-\omega_{\k_1})-
\mathcal{P}\frac{i}{\omega_{\k}-{\omega_{\k_1}}}\right\}.
\label{eq:gedmarkov}
\end{eqnarray}
The memory $G_{\omega_\k}^{\omega_{\k_1}}$ contains a Dirac delta distribution
with respect to the frequencies $\omega_\k$.
This allows us to derive from the PESC Eqs.~(7) 
closed equations for dynamical variables that depend only on the frequencies.
To this end, we define the following averaged quantities:
\numparts
\begin{eqnarray}
n^{\up/\down}_{\omega_1}&:=\frac{\sum_\k n^{\up/\down}_{\k}\delta(\omega_\k-\omega_1)}
{\sum_\k\delta(\omega_\k-\omega_1)}\\
\mbf s^{\perp}_{{\omega_{1}}}&:=\frac{\sum_\k \mbf s^{\perp}_{\k}\delta(\omega_\k-\omega_1)}
{\sum_\k\delta(\omega_\k-\omega_1)}.
\end{eqnarray}
\endnumparts
Due to the delta distribution in Eq.~(\ref{eq:gedmarkov}),
it becomes clear that the first term in Eq.~(\ref{eq:modmarkovA}) for the
spin-up and spin-down occupations disappears.
Therefore, performing the Markov limit of 
Eqs.~(7) 
and averaging over the angles, we obtain the following equations for the averaged variables
$n^{\up/\down}_{{\omega_{1}}}$ and $s^{\perp}_{\omega_{1}}:=|\mbf s^{\perp}_{{\omega_{1}}}|$:
\numparts
\begin{eqnarray}
\ddt n^\up_{\omega_1}=&cD({\omega_2})\bigg\{ b^+ n_{\omega_2}^\down - b^-n^\up_{\omega_1}
\label{eq:modmarkovdeltaA}
- 2b^0  n^\up_{\omega_1} n^\down_{\omega_2}\bigg\} \\
\label{eq:modmarkovdeltaB}
\ddt n^\down_{\omega_2}=&cD({\omega_1})\bigg\{ b^- n_{\omega_1}^\up - b^+n^\down_{\omega_2}
+ 2b^0 n^\up_{\omega_1}n^\down_{\omega_2}\bigg\} \\
\label{eq:modmarkovdeltaC}
\ddt s^\perp_{\omega_1}=&
-c \bigg\{\big(D(\omega_0)+D(\omega_2)\big)\frac{\langle {S^\perp}^2\rangle}2 +
D(\omega_1)\langle {S^\|}^2\rangle +\nn&
-\frac{\langle S^\|\rangle}2
\big(D(\omega_0)n^\up_{\omega_0}-D(\omega_2)n^\down_{\omega_2}\big)
\bigg\}s^\perp_{\omega_1},
\end{eqnarray}
\endnumparts
where $\omega_0:=\omega_1-\omega_M$ and
$\omega_{2}:=\omega_{1}+\omega_{M}$. Here, we have used that in the
quasi-continuum limit $\sum_\k \to \int d\omega D(\omega)$ 
with $D(\omega)$ being the density of states (DOS),
and thus:
\begin{eqnarray}
&\sum_\k\Re \big(G_{\omega_\k}^{\omega_{\k_1}}\big)=
\frac{\Jsd^2}{\hbar^2}\frac\nMn V \pi\sum_\k
\delta(\omega_\k-\omega_{\k_1})=
\underbrace{\frac{\Jsd^2}{\hbar^2}\frac\nMn V \pi}_{=:c} D(\omega_1).
\end{eqnarray}
Therefore, it can be seen from Eqs.~(\ref{eq:modmarkovdeltaA}-$b$) that in the Markov limit of 
the PESC equations, the only dynamical variables entering the equation of motion for
the spin-up electrons $n^\up_{\omega_1}$ with frequency $\omega_1$ are
$n^\down_{\omega_2}$ and $n^\up_{\omega_1}$ itself.
Equally, the time evolution of $n^\down_{\omega_2}$
only depends on $n^\up_{\omega_1}$ and $n^\down_{\omega_2}$, so that this pair 
of occupations is completely decoupled from the rest of the dynamical variables.
Furthermore, the total number of electrons in the pair of occupations 
$n^\up_{\omega_1}$ and $n^\down_{\omega_2}$ is conserved,
since from Eqs.~(\ref{eq:modmarkovdeltaA}-$b$) it follows:
\numparts
\begin{eqnarray}
\ddt z_{\omega_1}=&0, 
\label{eq:n_conserv}
\end{eqnarray}
where
\begin{eqnarray}
z_{\omega_1}:=&D(\omega_1)n_{\omega_1}^\up+D(\omega_2)n_{\omega_2}^\down.
\end{eqnarray}
\endnumparts

Eq.~(\ref{eq:n_conserv}) allows us to merge Eqs.~(\ref{eq:modmarkovdeltaA}-$b$) 
into a one-dimensional differential equation:
\begin{eqnarray}
\ddt x_{\omega_1}&=2 c b^0x_{\omega_1}^2-c\big[D(\omega_1)b^+ +D(\omega_2)b^-+2 b^0z_{\omega_{1}}\big]
x_{\omega_1} \nn& +c D(\omega_1)b^+ z_{\omega_1}
\label{eq:riccati}
\end{eqnarray}
for the spectral electron density in the spin-up subband
 $x_{\omega_1}:=D(\omega_1)n_{\omega_1}^\up$.

The last term in Eq.~(\ref{eq:modmarkovdeltaA}) and (\ref{eq:modmarkovdeltaB}), 
respectively, is a consequence of the source terms $\bstd^{I.2}$ in 
Ref.~\onlinecite{Cygorek:14_1} which 
were associated with Pauli blocking in the golden rule-type
rate equations that did not account for the precession of the correlations.
This fact is also visible here,
since for $n^\up_{\omega_1}\approx 1$, Eq.~(\ref{eq:modmarkovdeltaA}) yields
\numparts
\begin{eqnarray}
\label{eq:n1gleich1A}
\ddt n^\up_{\omega_1}&=
cD({\omega_2})\big(\langle {S^\perp}^2\rangle -\frac{\langle S^\|\rangle}2\big) 
\big( n^\down_{\omega_2}- 1\big),
\end{eqnarray}
while without the last term of Eq.~(\ref{eq:modmarkovdeltaA}), the limit would be
\begin{eqnarray}
\ddt n^\up_{\omega_1}&=
cD({\omega_2})\big(\langle {S^\perp}^2\rangle 
\big( n^\down_{\omega_2}- 1\big) + \frac{\langle S^\|\rangle}2 
\big( n^\down_{\omega_2}+1\big) \big) 
\label{eq:n1gleich1B}
\end{eqnarray}
\endnumparts
Since the minimal value of $\langle {S^\perp}^2\rangle$ is $\frac{35}{12}$ which is greater than the
maximal value of $\frac{\langle S^\|\rangle}2$ of $\frac 54$, the r.h.s. of
Eq.~(\ref{eq:n1gleich1A}) is always non-positive, so an over-occupation of $n^\up_{\omega_1}$
with values greater than $1$ is averted. In contrast, in Eq.~(\ref{eq:n1gleich1B}) 
the occupation $n^\up_{\omega_1}$ can exceed the physically reasonable limit of $1$, e.~g., in
the case $n^\down_{\omega_2}\approx 1$.
Thus, again, the terms resulting from $\bstd^{I.2}$ in Ref.~\onlinecite{Cygorek:14_1} 
are shown to provide for Pauli blocking effects.

The equation of motion (\ref{eq:modmarkovdeltaC}) for the electron spin 
component perpendicular to the Mn magnetization suggest an almost 
exponential decay to zero, but 
the occupations of the spin-up electrons at $\omega_1-\omega_M$ and 
spin-down electrons at $\omega_1+\omega_M$ enter in the effective decay rate.
The appearance of the occupations is also due to the $\bstd^{I.2}$ terms.
Here, they do not represent Pauli blocking, but are a remnant of the 
Landau-Lifshitz-Gilbert-like damping term structure in the Markov limit in 
Ref.~\onlinecite{Cygorek:14_1}, since:
\begin{eqnarray}
&\frac{\langle S^\|\rangle}2
 \big(D(\omega_0)n^\up_{\omega_0}-D(\omega_2)n^\down_{\omega_2}\big)
s^\perp_{\k_1}=
\frac{\langle S^\|\rangle}2\big(\frac{z_{\omega_0}-z_{\omega_1}}2+
s^\|_{\omega_1}\big)s^\perp_{\k_1}
\end{eqnarray}
and
\begin{eqnarray}
& \bigg[\mbf s_\omega \times \big(\mbf s_{\omega_1} \times 
\langle \mbf S \rangle\big) \bigg]_\perp=
\langle S^\|\rangle s^\|_\omega s^\perp_{\omega_1}.
\end{eqnarray}

In fact, comparing the derivation of the Markov limit with and without accounting
for the precession of the correlations it can easily be seen that the
PESC Markov Eqs.~(12) 
lead to the Markov Eq. (10) of
Ref.~\onlinecite{Cygorek:14_1}, when $\omega_M$ is set to zero in the 
memory terms $G_{\omega}^{\omega_1\pm\omega_M}$.

\subsection{Analytical solutions}  
The Markov limit of the PESC equations allows us to find analytic solutions
which we will derive in the following.
\subsubsection{without Pauli blocking}
If the terms resulting from ${b_{\beta\k_1}^{\alpha\k_2}}^{I.2}$ are neglected,
Eqs.~(\ref{eq:modmarkovdeltaA}) and (\ref{eq:modmarkovdeltaB}) yield:
\begin{eqnarray}
\ddt x_{\omega_1}&=-c\big(D(\omega_1)b^+ + D(\omega_2)b^-\big)x_{\omega_1}+ 
c D(\omega_1)b^+  z_{\omega_1}.
\label{eq:ohnePauli}
\end{eqnarray}
The solution of Eq.~(\ref{eq:ohnePauli}) decays exponentially:
\numparts
\begin{eqnarray}
x_{\omega_1}(t)&=
\big(x_{\omega_1}(0)-\xi_{\omega_{1}}\big)e^{-\eta_{\omega_{1}} t}+\xi_{\omega_{1}} ,
\end{eqnarray}
with
\begin{eqnarray}
\eta_{\omega_{1}}:&=c\big(D(\omega_1)b^+ + D(\omega_2)b^-\big) \\
\xi_{\omega_{1}}:&=\frac{D(\omega_1)b^+ }{D(\omega_1)b^+ + D(\omega_2)b^-}z_{\omega_1}
\end{eqnarray}
\endnumparts
It should be noted that 
for $D(\omega_2)\to D(\omega_1)$,
the rate $\eta_{\omega_{1}}$ 
reaches the same value as for the rate equations of 
Ref.~\onlinecite{Cygorek:14_1} and Fermi's golden 
rule\cite{Thurn:13_1,KossutRate3D} when only the parabolic band energy is
accounted for the initial and final states.
Here, $\eta_{\omega_{1}}$ describes rates that can be derived
with Fermi's golden rule,
when the mean field energy difference between electrons in the spin-up and 
spin-down subbands $\hbar\omega_M$ is substituted into the band structure and
transitions between these now non-degenerate subbands are considered.

Furthermore, without the terms originating from 
${b_{\beta\k_1}^{\alpha\k_2}}^{I.2}$, 
the perpendicular component of the electron spin changes according to
\begin{eqnarray}
\ddt s^\perp_{\omega_1}=&
-\underbrace{c \bigg\{\big(D(\omega_0)+D(\omega_2)\big)
\frac{\langle {S^\perp}^2\rangle}2 +D(\omega_1)\langle {S^\|}^2\rangle
\bigg\}}_{\gamma^\perp_{\omega_1}:=}s^\perp_{\omega_1}
\label{eq:sperp}
\end{eqnarray}
which is solved by
\begin{eqnarray}
s^\perp_{\omega_1}(t)=s^\perp_{\omega_1}(0)e^{-\gamma^\perp_{\omega_1}t}
\end{eqnarray}

It should be noted that neglecting the ${b_{\beta\k_1}^{\alpha\k_2}}^{I.2}$
terms in the rate equations 
without accounting for the precession of the correlations 
yielded the same expression for the rate
that can also be obtained by letting $D(\omega_0)$ and $D(\omega_2)$ go to 
$D(\omega_1)$ in Eq.~(\ref{eq:sperp}). In Ref.~\onlinecite{Cygorek:14_1} it was found
that including the precession of the correlation in the calculation did not 
significantly change the spin dynamics of the perpendicular component.
Now, this can be understood by Taylor-expanding the DOS. 
Since in three dimensions, the DOS is proportional to the square root of $\omega$,
we find:
\numparts
\begin{eqnarray}
D(\omega_1\pm \omega_M)&=D(\omega_1)\sqrt{1\pm \frac{\omega_M}{\omega_1} }
\label{eq:DDzweiteOrdnungA}
\end{eqnarray}
{and therefore}
\begin{eqnarray}
D(\omega_0)+D(\omega_2)&=2D(\omega_1)+\mathcal{O}
\left(\left(\frac{\omega_M}{\omega_1}\right)^2\right).
\label{eq:DDzweiteOrdnungB}
\end{eqnarray}
\endnumparts
Since the difference between the 
rates for the perpendicular component with and without 
accounting for the precession of the correlations is of second
order of the ratio $\frac{\omega_M}{\omega_1}$, 
significant deviations can only be expected for small values of $\omega_1$.
There, however, the DOS is rather small.

\subsubsection{with Pauli blocking}
Eq.~(\ref{eq:riccati}) is a Riccati differential equation with 
constant coefficients. 
Also in the case of golden rule-type rate equations derived from the 
original quantum kinetic theory,
where the precession of the correlations around the Mn magnetization is neglected,
we found an equation for the parallel electron spin component of this 
form\cite{Cygorek:14_1}. The solutions of Eq.~(\ref{eq:riccati}) can be obtained 
along the line of the Appendix in Ref.~\onlinecite{Cygorek:14_1}:
\numparts
\begin{eqnarray}
x_{\omega_1}(t)&=\frac{\mu_{\omega_{1}}}{2cb^0}-\frac {\nu_{\omega_{1}}}{2cb^0} 
\t{tanh}\left(\frac {\varphi_{\omega_{1}}}{ 2}+\nu_{\omega_{1}} t \right), 
\end{eqnarray}
with
\begin{eqnarray}
\mu_{\omega_{1}}&=\frac c2 \big[D(\omega_1)b^+ +D(\omega_2)b^-+2b^0z_{\omega_1}\big], \\
\nu_{\omega_{1}}&=\sqrt{\mu^2-2c^2D(\omega_1)b^+b^0  z_{\omega_1}},
\end{eqnarray}
\endnumparts
where $z_{\omega_1}$ 
and $\varphi_{\omega_{1}}$ are determined by the initial values of $n^\up_{\omega_1}$
and $n^\down_{\omega_2}$.

Finally, the time dependence of the perpendicular spin component can be calculated
using the analytical expressions for $n_{\omega_0}^\up$ and $n_{\omega_2}^\down$ 
and reads:
\begin{eqnarray}
s^\perp_{\omega_1}(t)&=s^\perp_{\omega_1}(0)e^{-\gamma_{\omega_1}^\perp t}
e^{-b^0 c \int\limits_0^t dt'
\big(D(\omega_0)n^\up_{\omega_0}(t')-D(\omega_2)n^\down_{\omega_2}(t')\big)}
\label{eq:senkr_int}
\end{eqnarray}
In order to explicitly give the corresponding analytical expressions we have to distinguish
two cases:\\ For $\omega_1<\omega_M$, $D(\omega_0)$ vanishes and we find from 
Eq.~(\ref{eq:senkr_int}):
\numparts
\begin{eqnarray}
s^\perp_{\omega_1}(t)&=s^\perp_{\omega_1}(0)e^{-\gamma_{\omega_1}^\perp t}
e^{(b^0c z_{\omega_1}-\frac 12 \mu_{\omega_1})t}
\sqrt{\frac{\t{cosh}\big(\frac{\varphi_{\omega_1}}2+\nu_{\omega_1}t\big)}{
\t{cosh}\big(\frac{\varphi_{\omega_1}}2\big)}},
\end{eqnarray}
and for $\omega_1>\omega_M$ we obtain:
\begin{eqnarray}
s^\perp_{\omega_1}(t)&=s^\perp_{\omega_1}(0)e^{-\gamma_{\omega_1}^\perp t}
e^{(b^0c z_{\omega_1}-\frac 12 \mu_{\omega_0}-\frac 12 \mu_{\omega_1})t}
\times\nn&\times
\sqrt{\frac{\t{cosh}\big(\frac{\varphi_{\omega_1}}2+\nu_{\omega_1}t\big)}{
\t{cosh}\big(\frac{\varphi_{\omega_1}}2\big)}}
\sqrt{\frac{\t{cosh}\big(\frac{\varphi_{\omega_0}}2+\nu_{\omega_0}t\big)}{
\t{cosh}\big(\frac{\varphi_{\omega_0}}2\big)}}.
\end{eqnarray}
\endnumparts

\subsection{Numerical studies}
\begin{figure}[h!]
\centering
\includegraphics[width=\textwidth]{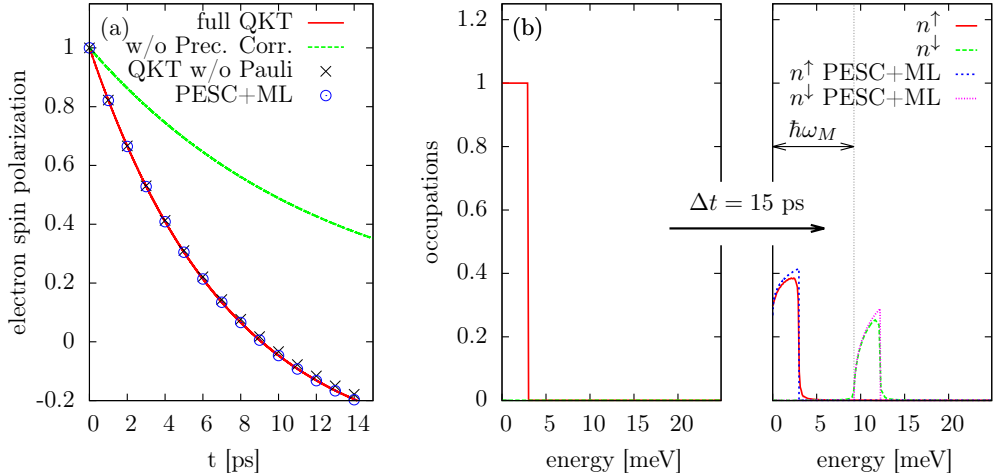}
\caption{(a): Time evolution of the electron spin polarization of 
an initially step-like electron occupation in the spin-up subband.
The red solid line is the result of a calculation using the full quantum kinetic theory (QKT) 
(cf. Ref.~(\onlinecite{Thurn:12})). The green dashed line is derived from 
the rate equations
without taking the precession of the correlations (w/o Prec. Corr.) 
into account (cf.
Ref.~(\onlinecite{Cygorek:14_1})). The black crosses represent calculations of the 
quantum kinetic theory
where the Pauli blocking terms ${b_{\beta\k_2}^{\alpha\k_1}}^{I.2}$ are neglected.
The blue circles describe the spin dynamics according to the Markov limit of 
the PESC Eqs.~(12). 
(b): Electron distributions at times $t=0$ ps and 
$t=15$ ps calculated using the full quantum kinetic theory and the 
Markov limit PESC equations, respectively.
}
\label{fig:einzelfermi}
\end{figure}

The validity of the approximations used to derive the 
Markov limit PESC Eqs.~(12) 
is now
checked by comparing the predicted spin dynamics with the results of the 
full quantum kinetic theory of Ref.~\onlinecite{Thurn:12} including also the residual terms
that are denoted as ${b^{\alpha\k_2}_{\beta\k_1}}^{Res}$ in Eq.~(\ref{eq:grundglC}). 
We modeled a bulk DMS of 
Zn$_{0.93}$Mn$_{0.07}$Se with the following parameters: the Kondo coupling
constant $\Jsd=12$ meVnm$^3$, the effective mass $m_e=0.21\;m_0$ and an 
initial average Mn spin of $\frac 12\hbar$ 
along the z-axis.
The difference between the different levels of theory, especially the role of the
Pauli blocking terms, can be particularly highlighted by choosing initial non-equilibrium conditions, 
where the initial electron occupations are modeled by step functions
(cf. Fig.~\ref{fig:einzelfermi}).
\footnote{If the Pauli 
blocking terms are neglected, the 
equations are linear in $\mbf s_{\k}$ and $n_\k$. Therefore, the solutions of the equations for
other initial occupations can be written as linear combinations of the solutions for the
step functions.} 

In a first calculation, the spin-up subband occupation was initially   
a step function with a cut-off energy at $\mu=3$ meV for electrons in the spin-up subband, 
while the spin-down subband was totally unoccupied for $t=0$. The results are shown in 
Fig.~\ref{fig:einzelfermi} (a) where we plot the modulus of the total spin polarization
\begin{eqnarray}
  \mbf s_{{tot}}=\bigg(\sum\limits_{\k}\mbf s_{\k}\bigg)\bigg(\sum_\k\frac 12 n_{\k}\bigg)^{-1}.
\end{eqnarray}
There, the spin polarization 
is shown to decrease almost exponentially 
from the initially completely polarized configuration to a negative value 
according to the full quantum kinetic theory. While the 
calculation without accounting for the precession of the 
correlations
deviates from the full quantum kinetic theory significantly, as it was already found in 
Ref.~\onlinecite{Cygorek:14_1}, the Markov limit of the PESC equations is able to 
reproduce the results of the full quantum kinetic theory almost perfectly. By comparison with 
the calculation neglecting the source terms 
${b_{\beta\k_2}^{\alpha\k_1}}^{I.2}$ it can be seen that for the initial values
used in this calculation, Pauli blocking effects are of minor importance.

Fig.~\ref{fig:einzelfermi} (b) depicts the energetic redistribution of the electrons:
The initial step-like spin-up occupation evolves into a structure with
two peaks; one in the spin-down and one in the spin-up band. The spin-up electrons
with energy $\hbar\omega_1$ are redistributed to states with energy 
$\hbar\omega_1+\hbar\omega_M$ which is predicted by the Markov limit of the 
PESC Eqs.~(7) 
due to terms proportional to $\delta(\omega_\k
-(\omega_{\k_1}\pm\omega_M))$. 
In contrast, 
when the precession of the correlations are neglected,
the spin-down peak builds up at the same energetic position as the spin-up peak
as the shift by  $\hbar\omega_{M}$, which accounts for the precession-like dynamics of the
Mn-carrier correlations,  is missing in the delta distribution.
The skewness of the peaks in 
Fig.~\ref{fig:einzelfermi} (b) arises from the square-root dependence of the DOS
on the energy in a three dimensional system. 
The fact that a small tail is found below the spin-down peak 
representing occupations of states with energies
lower than $\hbar\omega_M$ as well as a build-up of a high energy tail of the 
spin-up peak demonstrate slight non-Markovian deviations from the 
Dirac delta-like memory in the Markov limit of Eq.~(\ref{eq:gedmarkov}). 
These effects are, however, too small to influence the 
time dependence of the total spin polarization significantly. 

\begin{figure}[t]
\centering
\includegraphics[width=\textwidth]{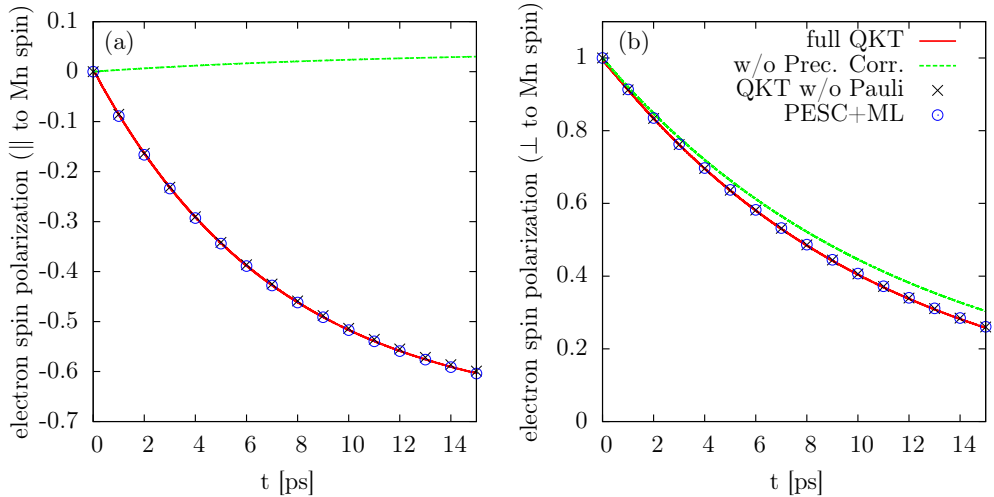} 
\caption{Time evolution of the spin polarization parallel (a) and
perpendicular (b) to the Mn magnetization. The initial electron distribution 
is the same as in Fig.~\ref{fig:einzelfermi}, except that the Mn magnetization was
rotated $90^\circ$ into a direction perpendicular to the initial electron spin
(key as in Fig.~\ref{fig:einzelfermi}a).}
\label{fig:senkrecht}
\end{figure}

Fig.~\ref{fig:senkrecht} displays the time evolution of the electron spin 
polarization for a situation
where the initial conditions were chosen as in the calculations
for Fig.~\ref{fig:einzelfermi}, except that the average Mn spin is now 
turned 90$^\circ$ 
away from the electron spin. Unlike the case discussed before, 
the spin polarization vector $\mbf s_{tot}$ now has components
parallel and perpendicular to the Mn magnetization. 
Fig.~\ref{fig:senkrecht} (a) shows a build-up of the
spin polarization parallel to the Mn magnetization 
according to the full quantum kinetic theory which coincides with the calculations
in the Markov limit of the PESC equations and the simulations without accounting
for Pauli blocking. As in the previous calculations, the 
solution of the golden rule-type rate equation of 
Ref.~\onlinecite{Cygorek:14_1} 
deviates significantly from the other calculations, since the energetic 
redistribution of the electrons is neglected.
The time evolution of the perpendicular electron spin shown in 
Fig.~\ref{fig:senkrecht} (b), however, is almost the same in all of the above 
calculations which can be understood by considering Eqs.~(23).

\begin{figure}[t!]
\centering
\includegraphics[width=\textwidth]{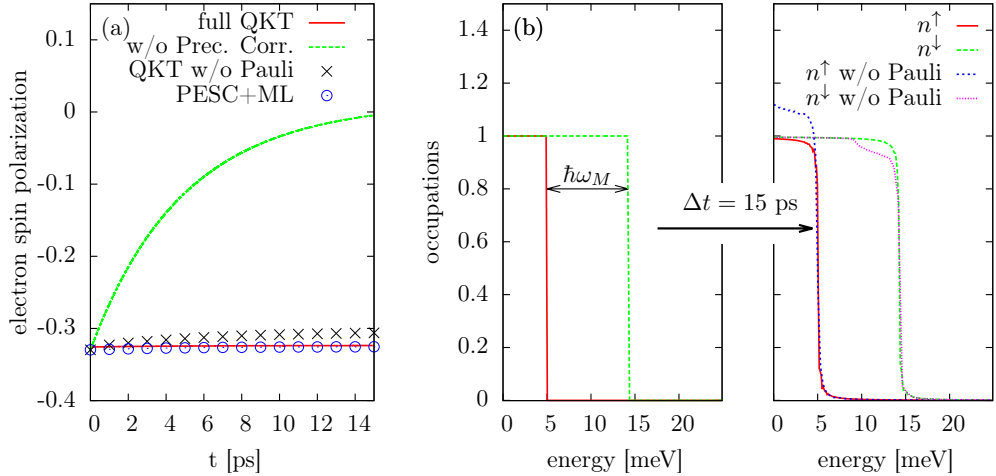}
\caption{Time evolution of the spin polarization (a) and 
electron redistribution (b) of an initially step-like
occupation of electrons where the difference of the cut-offs of the
spin-up and spin-down subband occupations is $\hbar\omega_M$. 
In (b), the occupations are plotted for calculations based on the full 
quantum kinetic theory with and without the terms
${b_{\beta\k_2}^{\alpha\k_1}}^{I.2}$
accounting for Pauli blocking.}
\label{fig:doppelfermi}
\end{figure}

In the following, we want to discuss the effects of Pauli blocking. To this end, 
we study a case, where both subbands are initially partly occupied
and where the spin dynamics is especially clear: 
We use initial conditions that describe a situation, where 
the spin polarization is expected to be nearly constant if Pauli blocking is taken into account.
Then, if we calculate the spin dynamics while neglecting the terms responsible for 
Pauli blocking effects, we can attribute the non-constant behavior to these effects.
If the Hamiltonian~(\ref{eq:kondoHam}) is treated on the mean field level, 
the equilibrium occupations at $T=0$ of the spin-down and spin-up subbands follow spin-split 
Fermi distributions, i.~e., step-functions, whose cut-off energies, 
measured from the respective band edge, differ by $\hbar\omega_M$.  
The results of the calculations with these initial conditions are given in Fig.~\ref{fig:doppelfermi}.
Since the full quantum kinetic theory also accounts for the correlation energies,
very small changes in the electron spin polarization are found, which are due to
the build-up of correlations over the course of time. 
Fig.~\ref{fig:doppelfermi}~(a) shows that the Markov limit of the PESC equations 
again yields nearly the same spin dynamics as the full quantum kinetic theory, 
while 
the rate equations without precession of the correlations  
fail to describe the spin polarization, since
different states are coupled due to the neglect of $\omega_M$ in the Dirac 
delta function of the memory. The calculations involving the quantum kinetic theory without the
Pauli blocking terms lead to surprisingly small deviations 
from the full quantum kinetic theory concerning 
the total spin polarization displayed in Fig.~\ref{fig:doppelfermi} (a),
despite the unphysical build-up of occupations $n^\up_{\omega}>1$ for some values of
$\omega$ seen in Fig.~\ref{fig:doppelfermi} (b).
The electron occupations after $t=15$ ps depicted in Fig.~\ref{fig:doppelfermi} (b) 
essentially follow the original step functions for the two spin orientations, 
but since the edges are smoothed
due to the non-Markovian deviations from the delta distribution in the memory,
they resemble Fermi distributions with an finite effective temperature. 

\section{Conclusion}
We have derived effective equations of motion [cf.~Eqs.~(7)]
for the conduction band electron spins and
occupations starting from a microscopic quantum kinetic theory using a 
rotating-wave-like Ansatz. 
These equations  account for the precession of 
the electron spins around the effective magnetic field due to the Mn magnetization
as well as for a precession-like dynamics 
of the electron-Mn correlation functions. 
Therefore, in this article they are referred to  
as {\em precession of electron spins and correlations} (PESC) equations. 
The PESC  equations can be more easily interpreted as the original 
quantum kinetic equations. They also provide an important speed-up of the numerics,
in particular, when the Markov limit of the PESC equations is used. 
The spin dynamics for high Mn doping in three dimensional systems derived 
from our effective equations in the Markov limit are demonstrated 
by numerical calculations 
to agree well with the corresponding results of the original quantum kinetic theory.
This  resolves the deficiency of the 
golden rule-type rate equations of Ref.~\onlinecite{Cygorek:14_1}
for the  case of a finite initial impurity magnetization. 
Even though the PESC equations can in principle describe
non-Markovian effects as well as Pauli blocking, the numerical studies suggest
that these are of minor importance for the time evolution of the total 
electron spin polarization, at least for the situations studied in this paper.

The Markov limit of the PESC Eqs.~(12) 
can be readily interpreted:
For a positive coupling constant $\Jsd$, spin-up electrons can gain energy 
in the mean field due to the Mn magnetization by
a spin-flip process to the spin-down subband. Due to the total energy conservation,
this energy is transformed into kinetic energy. Therefore, a spin-up state
couples effectively only to a spin-down state with 
a kinetic energy $\hbar\omega_M$ greater than the spin-up state energy and 
vice-versa. The resulting equation can be solved analytically yielding 
a time dependence of the electron spin polarization following a tanh-function.
If the Pauli blocking terms are neglected, the equations are solved by a simple
decaying exponential function with a rate that can also be obtained by applying Fermi's 
golden rule, if the mean field energy of an electron in the effective field of the
Mn magnetization is included into the single particle energies.

Numerical studies of the energetic redistribution of the electrons in the full
quantum kinetic theory support the findings of the delta-like coupling of states in energy space
in general, but slight deviations from this Markovian prediction can be seen
especially in the smoothing of sharp edges of the initial electron occupations.
It is expected that the non-Markovian features will be more important in 
two-dimensional systems\cite{Thurn:13_1,Thurn:13_2}. 
The PESC equations in the Markov limit derived in this paper can provide a suitable framework for further 
investigations of these effects. In addition, their numerical simplicity allows for the
introduction of other mechanisms of spin relaxation to study reliably their competition with
the $s$-$d$ exchange interaction which would be a demanding task within
the original quantum kinetic theory.

\ack
We acknowledge the support by the Deutsche Forschungsgemeinschaft
through the Grant No. AX 17/9-1.


\providecommand{\newblock}{}

\end{document}